\documentclass[reprint,amsmath,amssymb,aps,prl]{revtex4-1}

\usepackage{graphicx}
\usepackage{dcolumn}
\usepackage{bm}
\usepackage{hyperref}
\usepackage[mathlines]{lineno}

\begin{document}

\title{High-Frequency Gravitational-Wave Detection Using a Chiral Resonant Mechanical Element and a Short Unstable Optical Cavity}

\author{Yi Chen$^{1}$}
\email{yi.chen@partner.kit.edu;chenyi221@gmail.com}
\author{Muamer Kadic$^{2,3}$}
\author{David E. Kaplan$^{4}$}
\author{Surjeet Rajendran$^{4}$}
\author{Alexander O. Sushkov$^{5}$}
\author{Martin Wegener$^{1,2}$}

\affiliation{$^{1}$Institute of Applied Physics, Karlsruhe Institute of Technology (KIT), 76128 Karlsruhe, Germany}

\affiliation{$^{2}$Institute of Nanotechnology, Karlsruhe Institute of Technology (KIT), 76021 Karlsruhe, Germany}

\affiliation{$^{3}$Institut FEMTO-ST, UMR 6174, CNRS, Universit\'{e} de Bourgogne Franche-Comt\'{e}, 25000 Besan\c{c}on, France}

\affiliation{$^{4}$Department of Physics $\&$ Astronomy, Johns Hopkins University, Baltimore, MD 21218, USA}

\affiliation{$^{5}$Department of Physics, Boston University, Boston, MA 02215, USA}

\date{\today}

\begin{abstract}
Present gravitational wave detectors are based on the measurement of linear displacement in stable optical cavities. Here, we instead suggest the measurement of the twist of a chiral mechanical element induced by a gravitational wave. The induced twist rotates a flat optical mirror on top of this chiral element, leading to the deflection of an incident laser beam. This angle change is enhanced by multiple bounces of light between the rotating mirror and an originally parallel nearby fixed flat mirror. Based on detailed continuum-mechanics calculations, we present a feasible design for the chiral mechanical element including the rotating mirror. Our approach is most useful for signals in the frequency band $1$ -- $100\,\rm kHz$ where we show that fundamental metrological limits would allow for smaller shot noise in this setup in comparison to the detection of linear displacement. We estimate a gravitational wave strain sensitivity between $10^{-21}/\sqrt{\rm{Hz}}$ and $10^{-23}/\sqrt{\rm{Hz}}$ at around $10\,\rm{kHz}$ frequency. When appropriately scaling the involved geometrical parameters, the strain sensitivity is proportional to frequency. 
\end{abstract}


\maketitle

The discovery of gravitational waves \cite{einstein1916approximative} by LIGO \cite{Abbott2016Obser, GW1512262016Abbott, GW1708172017Abbott, GW1701042017Abbott} has opened a new window into the universe. LIGO attains its peak sensitivity  around $200\,\rm{Hz}$  \cite{aasi2015advanced}, enabling it to detect the mergers of binary black holes that have masses $\approx 10 M_{\odot}$. There is a strong physics case to probe gravitational waves at higher frequencies ($10$ -- $100\,\rm kHz$) where one might observe mergers of smaller  ($\approx 1 M_{\odot}$) black holes, while also being able to search for the full spectrum \cite{Kaplan:2018dqx} of quasinormal modes from the ringdown of a merger observed in LIGO. In addition to these standard astrophysical signatures, detectors in this frequency band will also be sensitive to gravitational waves produced from the high temperature ($\approxeq $ PeV) universe and from new physics such as super-radiant bosonic clouds around black holes \cite{Arvanitaki:2009fg}. Simultaneously, these detectors can search for a variety of ultra-light dark matter \cite{Gambhir2016Zeptonewton, Graham:2015ifn, Arvanitaki:2016fyj}.

The fundamental difficulty in probing these higher frequencies arises from the fact that the shorter period reduces the time available for the signal to cause a measurable change (i.e., beat shot noise) in an experiment, diminishing the sensitivity of a LIGO-style stable optical cavity at these frequencies. In this Letter, we point out that the shot noise limit on sensitivity can be improved by measuring angular displacement in an unstable optical cavity as opposed to measuring linear displacement in a stable cavity. This sensitivity gain can only be attained if other sources of noise (such as thermal noise) are suppressed and if the quality factor of the cavity is sufficiently large. Due to the rapid scaling of event rates in gravitational wave detectors with sensitivity, even modest improvements in sensitivity lead to significant scientific payoff.

Our idea starts from converting an axial gravitational strain into a twist or rotation, which has been discussed in the context of chiral mechanical metamaterials\cite{Frenzel2017, fernandez2019new, chen2020mapping}. As shown in Fig.\,1, the detection principle of our detector is to measure the gravitational strain by monitoring the angle of a laser beam that is reflected by a mirror that is rotated by the induced twist. The strain-to-twist conversion factor, $K$, can be boosted in the vicinity of a high-quality-factor mechanical resonance. As illustrated in Fig.\,1(b), the magnitude of the deflection angle can be increased by multiple round trips of light between two flat mirrors, $N$ \cite{Hogan_2011}. To reduce shot noise by reducing the diffraction limit on measuring this angle, the diameter of the beam and hence the size of the rotating mirror need to be as large as possible. However, large size corresponds to large mass, which reduces the resonance frequency unless the torsional stiffness of the chiral mechanical element can be made very large. Furthermore, the chiral torsional mechanical eigenmode should ideally correspond to the lowest mechanical eigenfrequency of the overall setup. Otherwise, deformations other than a pure rotation of the mirror may become increasingly important. For example, a warping of the previously flat mirror surface would lead to an unwanted distortion of the laser beam profile. Altogether, this means that the design of the resonant chiral mechanical element results from a non-trivial trade-off.

To investigate the elastic behavior of the chiral mechanical element shown in Fig.\,1(a) under the influence of a gravitational wave impinging along the $y$-direction, we solve Newton’s second law combined with Hooke’s law, cast into linear Cauchy continuum mechanics for the position and time-dependent displacement vector $\mathbf{u}=\mathbf{u}(\mathbf{r},t)$ given by
\begin{equation} \label{(1)} 
\rho \frac{\partial^2\mathbf{u}}{\partial t^2}=\mathbf{\nabla}\cdot{\bm \sigma}+\rho \mathbf{a}_{\rm gw}.
\end{equation} 
$\rho=\rho(\mathbf{r})$ is the local mass density, $\bm{\sigma}=\bm{\sigma}(\mathbf{r})=\mathbf{C}:\bm{\epsilon}$ the rank-2 stress tensor, given by the contraction '':'' of the rank-4 Cauchy elasticity tensor $\mathbf{C}=\mathbf{C}(\mathbf{r})$ and the rank-2 strain tensor $\bm{\epsilon}=\bm{\epsilon}(\mathbf{r},t)=\mathbf{\nabla}\mathbf{u}$, and $\mathbf{a}_{\rm gw}=\mathbf{a}_{\rm gw}(\mathbf{r},t)$ is the gravitational-wave acceleration vector. $\rho \mathbf{a}_{\rm gw}$ represents a body-force density \cite{misner1973gravitation} and is given by
\begin{equation} \label{(2)} 
\mathbf{a}_{\rm gw}=\frac{\partial^2}{\partial t^2}\mathbf{h}\cdot(\mathbf{r}-\mathbf{r}_0),
\end{equation}
with the gravitational-wave strain matrix $\mathbf{h}$ at angular frequency $\omega_{\rm gw}=2 \pi f_{\rm gw}$
\begin{equation} \label{(3)} 
\mathbf{h}=h_{\rm 0}\,{\rm cos}(k_{\rm gw} y-\omega_{\rm gw} t ) \left(\begin{array}{ccc} {-1} & {0} & {0}\\{0} & {0} & {0}\\{0} & {0} & {1} \end{array}\right) .
\end{equation}
The reference coordinate $\mathbf{r}_0$ can be chosen arbitrarily, e.g., as the center of mass of the setup, which includes a large cuboid plate underneath the chiral element shown in Fig.\,1(a). $h_0$ is the dimensionless strain amplitude of the gravitational wave. The dispersion relation is $\omega_{\rm gw} / k_{\rm gw}=c_0$, with the vacuum speed of light $c_0=3\times10^8\,{\rm m/s}$. We approximate the wave number as $k_{\rm gw}=0$, which is justified if the wavelength of the gravitational wave, $\lambda_{\rm gw}=2\pi / k_{\rm gw}$, is large compared to the size of the setup shown in Fig.\,1. For the frequency $\omega_{\rm gw}/(2\pi)=10\,{\rm kHz}$ chosen below, we have $\lambda_{\rm gw}=30\,{\rm km}$ and this approximation is well justified. The above equations are solved numerically by standard frequency-domain finite-element calculations using the software package Comsol Multiphysics \cite{thomas2000finite} and its MUMPS solver. The geometrical parameters are given in Fig.\,1. For the material parameters, we choose an isotropic (polycrystalline) version of diamond \cite{Ashby2011}, with a Young’s modulus of $E = 1.13 \times 10^3\,{\rm GPa}$, a loss tangent of ${\rm tan}(\delta) = 3 \times 10^{-6}$, a Poisson’s ratio of $\nu = 0.2$, and a mass density of $\rho = 3.510 \times 10^3\,{\rm kg/m^3}$. Diamond serves as a benchmark in the sense that it exhibits the largest transverse and longitudinal phonon phase velocities of all known natural materials. In essence, the phonon phase velocities combined with the geometrical parameters determine the mechanical eigenfrequencies of the system.

\begin{figure}[ht]
\includegraphics[width=5.3cm,angle=0]{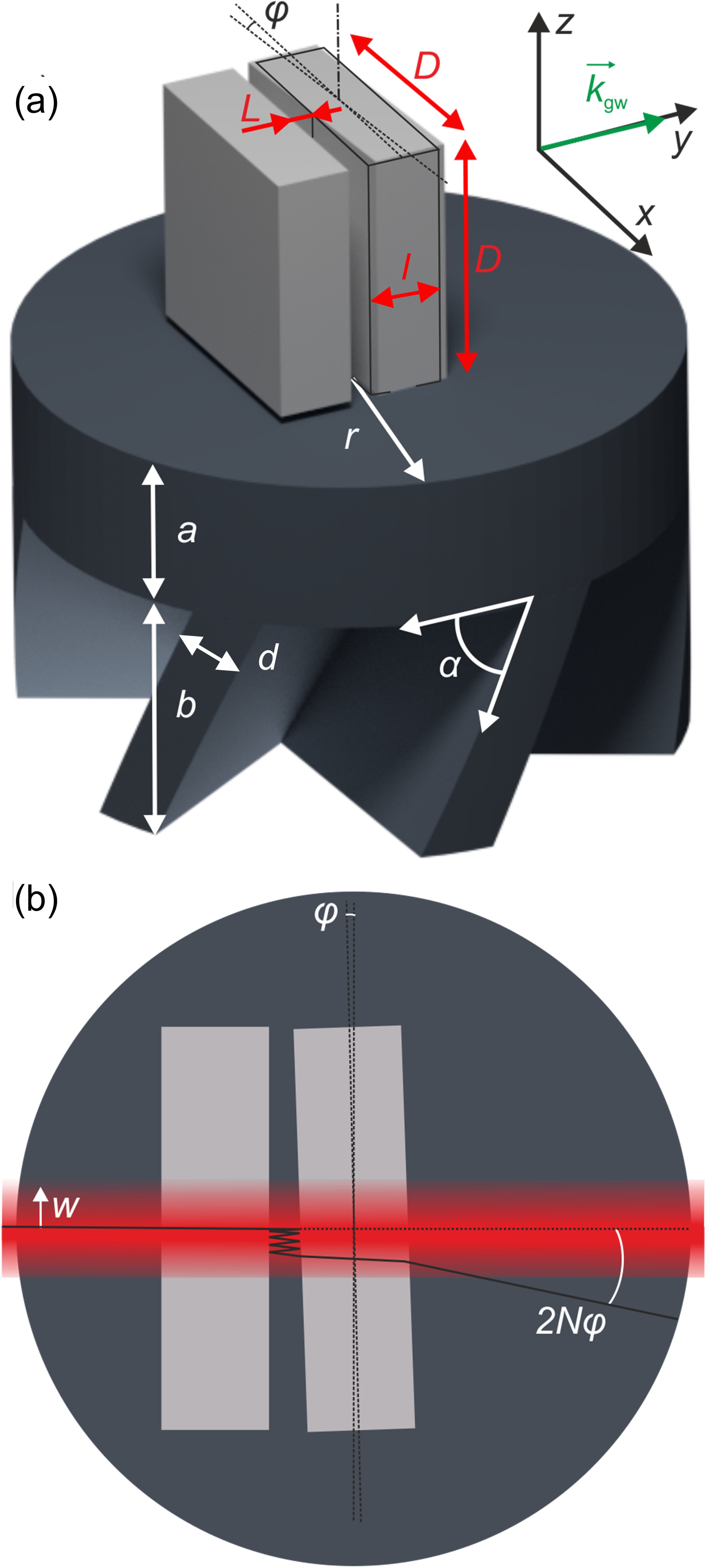}
\caption{(a) Illustration of the designed chiral mechanical element, leading to the resonant mechanical conversion of a local gravitational-wave induced strain amplitude $h_0$ into a twist, hence into a rotation angle amplitude $\varphi_0 = K h_0$ of the flat mirror substrate on top. The mirror size is $D$. Further geometrical parameters as well as the incident wave vector of the gravitational wave $\mathbf{k}_{\rm gw} = (0, k_{\rm gw}, 0)$ at frequency $f_{\rm gw} = k_{\rm gw} c_0/(2 \pi)$ are indicated. We choose $D = 5\,{\rm cm}$, $r = 6.5\,{\rm cm}$, $l = 1.5\,{\rm cm}$, $a = 3.0\,{\rm cm}$, $b = 5.0\,{\rm cm}$, $d = 1.3\,{\rm cm}$, and $\alpha = 65^\circ$. The shown structure is located on a large plate with volume $65\,{\rm cm} \times 65\,{\rm cm} \times 26\,{\rm cm}$. (b) Together with a second parallel flat optical mirror (the fixed holder of which is not depicted), the rotating mirror forms an unstable optical cavity with length $L = 1\,{\rm mm}$. The other sides of the two mirrors are anti-reflection coated. This optical cavity translates the rotation angle $\varphi_0$ into the deflection of an incident Gaussian beam with $1/e^2$ intensity radius $w = 5\,{\rm mm}$ by a deflection angle $2 N \varphi_0$. Here, $N$ is the mean number of round trips of light in the cavity. For clarity, the beam deflection effect is largely exaggerated and the angles are not to scale.}
\label{Figure1}
\end{figure}

The system in Fig.\,1 reacts with a time-harmonic twist angle $\varphi(t) = \varphi_0 {\rm cos}(\omega_{\rm gw} t +\phi_{\varphi})$ and a time-harmonic axial strain $\epsilon_{zz}(t) = \epsilon^{0}_{zz} {\rm cos}(\omega_{\rm gw} t +\phi_{\epsilon})$. An example response function of the twist angle amplitude $\phi_0$ and the axial strain amplitude $\epsilon^0_{zz}$ versus frequency $f_{\rm gw}=\omega_{\rm gw}/(2\pi)$ is depicted in Fig. 2. Its vertical axis is normalized with respect to the dimensionless gravitational-wave strain amplitude $h_0$. The axial strain amplitude $\epsilon_{zz}^0$ is dimensionless, the twist angle amplitude $\varphi_0$ is in units of radians. On resonance, it reaches a maximum of $\varphi_0>10^5 h_0$. For a mirror mount with a size of $D = 5\,{\rm cm}$ in Fig.\,1, the lowest eigenfrequency of the overall setup lies at around $10.4\,{\rm kHz}$. It is clear that working close to a mechanical eigenfrequency means that the bandwidth of the gravitational wave detector becomes small. Below the resonance frequency, the response increases $\propto f_{\rm gw}^2 \propto \omega_{\rm gw}^2$ because the body-force density $\rho \vec{a}_{\rm gw}$ is proportional to the second temporal derivative of ${\rm cos}(\omega_{\rm gw} t)$. To provide a broader overview of the behavior, the six lowest-frequency eigenmodes of the system are shown in Fig.\,S1 \cite{PRLSupp}.

\begin{figure}[ht]
\includegraphics[width=8.5cm,angle=0]{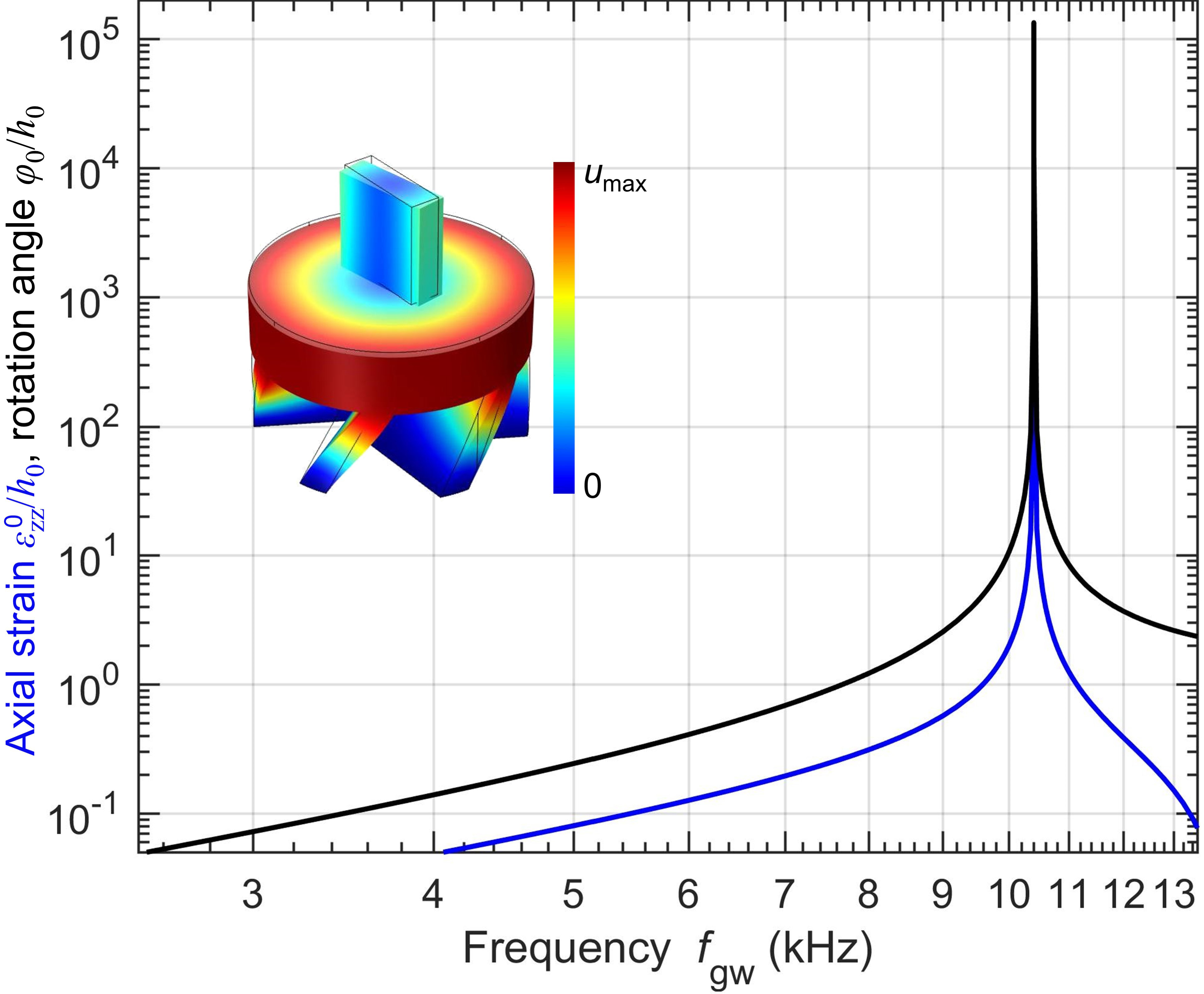}
\caption{Calculated response (logarithmic scale) of the axial strain amplitude $\epsilon_{\rm zz}^0$ (dimensionless) and the rotation angle amplitude $\varphi_0$ (in units of radians) of the mirror shown in Fig.\,1(a) versus gravitational-wave frequency $f_{\rm gw}$. Both quantities are normalized to the dimensionless gravitational-wave strain amplitude $h_0$. The resonant peak value of the rotation angle is $\varphi_0 > 10^5 h_0$, leading to, e.g., $\varphi_0 > K h_0$ with $K = 100$ over a bandwidth of about $100\,{\rm Hz}$. The inset exhibits the lowest eigenmode of the system shown in Fig.\,1. The false-color scale corresponds to the modulus of the displacement vector $\mathbf{u}$. Further eigenmodes are shown in Fig.\,S1 \cite{PRLSupp}.}
\label{Figure2}
\end{figure}

The example shown in Fig.\,2 can easily be scaled to other sizes provided that all aspect ratios are fixed and that the material parameters are fixed. For example, when increasing (decreasing) all geometrical parameters given in Fig.\,1 by a factor of ten, the frequencies on the horizontal axis of Fig.\,2 need to be scaled down (up) by a factor of ten, while the vertical axis remains unchanged.

Results for hybrid materials choices are shown in Fig.\,3. For example, when replacing the diamond mirror substrate at the top by silica ($\rm{SiO}_2$), while keeping the diamond bottom, unwanted deformations of the mirror become much more prominent as the mirror can no longer be approximately considered as a rigid body (c.f. Fig.\,S2 \cite{PRLSupp}). When going to the reverse hybrid structure of silica bottom and diamond top, the mirror rotates rigidly but the twist eigenfrequency decreases. On the other hand, hybrid architectures containing materials choices other than diamond are more accessible experimentally in terms of manufacturing and cost. Refer to Fig.\,S3 for other material combinations \cite{PRLSupp}.

Next, we estimate the gravitational-wave strain sensitivity achievable with the calculated (frequency-dependent) rotation angle amplitude
\begin{equation} \label{(4)} 
\varphi_0 = K h_0,
\end{equation}
with the frequency dependent strain-to-twist conversion factor, $K = K(f_{\rm gw})$. We consider an optical cavity of length $L$ composed of the rotating mirror and a second fixed mirror of the same size $D$ as shown in Fig.\,1(b). The two mirrors are flat and originally parallel, representing an unstable optical resonator on the level of wave optics \cite{koechner2013solid}. A Gaussian laser beam with a $1/e^2$ intensity radius $w$ impinges under normal incidence and is reflected by the mirrors $2N$ times, with the mean number of round trips of light in the cavity, $N$. In geometrical optics and for rotation angle $\varphi(t)$, normal incidence of light leads to a mean deflection angle of $2 N \varphi (t)$ (as illustrated in Fig.\,1(b)). Normal incidence of light with a sufficiently small error is important: To avoid that the beam walks off the edge of the mirror, the incidence angle $\beta$ of the laser beam with respect to the mirror surface normal must obey the condition $2LN|{\rm tan}(\beta)| \approx 2LN|\beta|\ll D$. For $D = 5\,{\rm cm}$, $L = 1\,{\rm mm}$, and $N = 2 \times 10^4$ (see below), this condition leads to $|\beta| \ll 1.3 \times 10^{-3}$ (i.e., $\ll 0.07$ degrees), which appears feasible. The parallelism of the two flat mirrors must be aligned with a comparable precision. 

To achieve a large mean deflection angle $2N\varphi_0$, large integers $N$ are clearly desirable. However, the product $NL$ is bounded by the fact that the dwell time of light between the two mirrors must not be larger than half of the gravitational wave temporal period $T_{\rm gw}/2=1/(2f_{\rm gw})$. Otherwise, the sign of the gravitational strain flips during the accumulation and the accumulated angle decreases again. This condition leads to the dwell-time inequality
\begin{equation} \label{(5)} 
NL \le \frac{c_0}{2f_{\rm gw}}=\frac{\lambda_{\rm gw}}{2}.
\end{equation}
For example, for $f_{\rm gw} = 10\,{\rm kHz}$, we obtain $NL \le 15\,{\rm km}$. With, e.g., $N = 2 \times 10^4$ this leads to $L \le 75\,{\rm cm}$. 

An independent bound for the product $NL$ arises from the fact that an incident Gaussian beam focused to an $1/e^2$ intensity radius $w$ at free-space optical wavelength $\lambda$ unavoidably diverges by diffraction of light when propagating over distance $y = NL$. However, this divergence is negligibly small provided that $NL$ is much smaller than the Rayleigh range $y_{\rm R}$ \cite{koechner2013solid}, i.e.,
\begin{equation} \label{(6)} 
NL \ll y_{\rm R} = \frac{\pi w^2}{\lambda}.
\end{equation}
For example, for $w = 5\,{\rm mm}$ and $\lambda = 1.064\,{\rm \mu m}$, we get $y_{\rm R} = 73.8\,{\rm m}$. With $N = 2 \times 10^4$, this leads to $L \ll 3.7\,{\rm mm}$. Clearly, for these parameters, this bound on $L$ is more than two orders of magnitude more stringent than the above dwell-time bound. For our choice of $L = 1\,{\rm mm}$ and $f_{\rm gw} = 10\,{\rm kHz}$, the dwell-time inequality is fulfilled as long as $N \ll 1.5 \times 10^7$. This means that the deflection angle $2N\varphi(t)$ well approximates the instantaneous gravitational wave strain for $N = 2 \times 10^4 \ll 1.5 \times 10^7$.

Achieving a value of  $N = 2 \times 10^4$ requires sufficiently high optical mirror reflectivity by using high-quality dielectric Bragg stacks. It also requires that the mirror size $D$ is sufficiently large compared to the $1/e^2$ intensity Gaussian beam radius $w = 5\,{\rm mm}$ to avoid cut-off losses. For example, for $D = 10 w = 5\,{\rm cm}$, the Gaussian intensity profile on the edge of the mirror decreases to $1/e^{10} \approx 4.5 \times 10^{-5}$ of the center value (cf. Fig.\,1(b)). 

\begin{figure}[ht]
\includegraphics[width=8.5cm,angle=0]{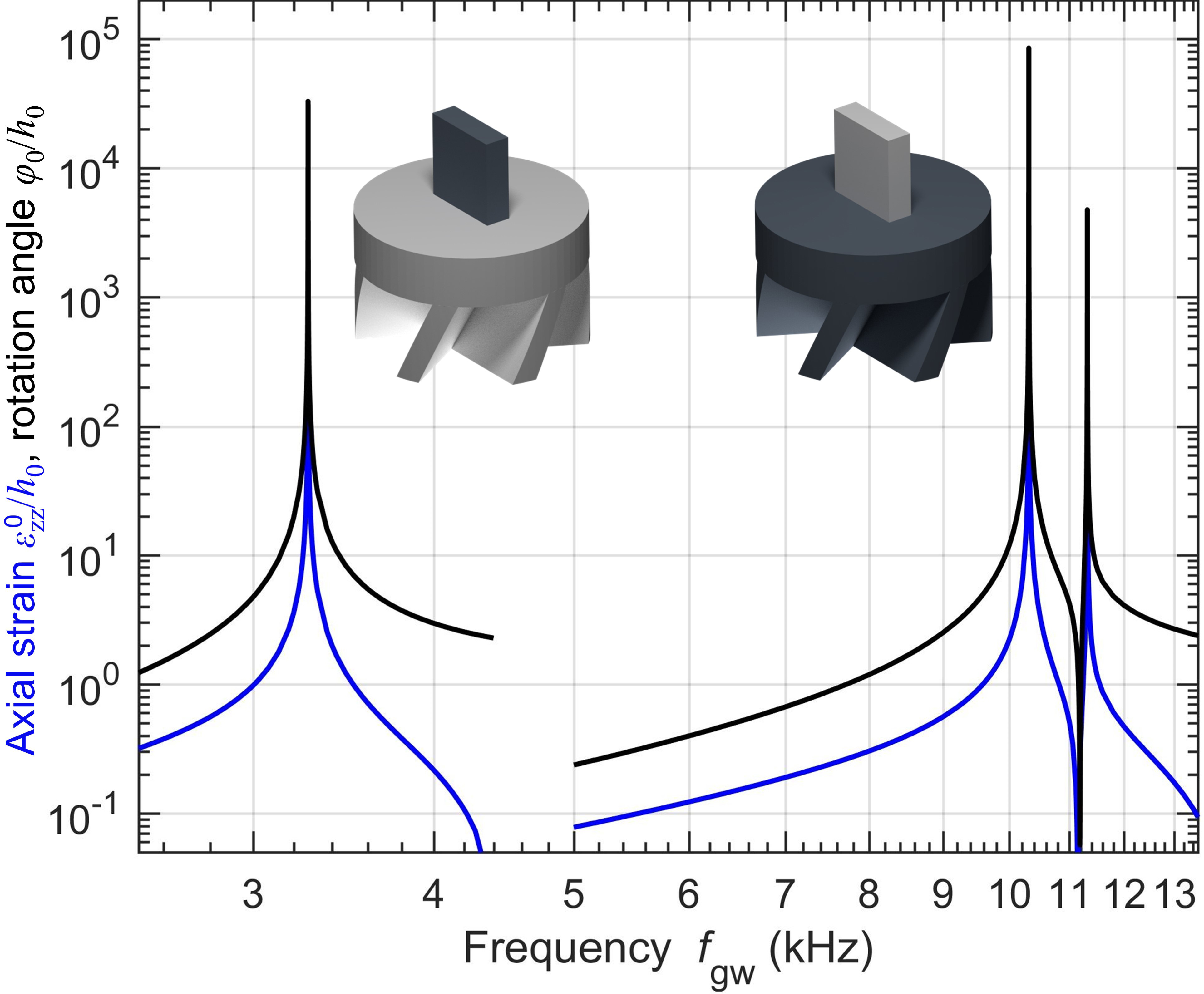}
\caption{Calculations as in Fig.\,2, but for hybrid silica-diamond architectures rather than the monolithic diamond architecture in Fig.\,2. Left: silica bottom, diamond mirror substrate. Right: diamond bottom, silica mirror substrate. Results for other material combinations are shown in Fig.\,S3 \cite{PRLSupp}.}
\label{Figure3}
\end{figure}

To measure the angular deflection, the emerging Gaussian laser beam could be focused \cite{Fundamentals2019Bahaa} by a lens of focal length $F$ to a Gaussian spot of $1/e^2$ intensity focus radius
\begin{equation} \label{(7)} 
w’ = \frac{\lambda F}{\pi w}.
\end{equation}
This focus radius is connected to the standard deviation, $s$, of the focus position via $s = w' / \sqrt{2}$. The localization error $\sigma$ of the focus position (or standard deviation of the mean value) for the rate of photons emerging from the cavity, $R$, is given by
\begin{equation} \label{(8)} 
\sigma = \frac{s}{\sqrt R}.
\end{equation}
We assume that we operate on a Fabry-Perot resonance, where the optical transmittance is unity and $R$ equals the rate of photons of the incident laser.
Finally, the focus displacement, $d$, due to the mean deflection angle $2 N \varphi_0$ is given by $d = F 2 N \varphi_0$. The minimum detectable gravitational strain amplitude $h_0$ corresponds to $d = \sigma$. Insertion of the above equations leads to the gravitational-wave strain sensitivity
\begin{equation} \label{(9)} 
h_0 = \frac{\varphi_0}{K}= \frac{1}{2 \pi \sqrt{2}}\frac{1}{K} \frac{\lambda}{N w} \frac{1}{\sqrt{R}} .
\end{equation}
Here, the focal length $F$ has dropped out, which means that one can choose an experimentally convenient focal length for the displacement measurement. 

The estimate (9) contains two aspects that are conceptually different from mechanical off-resonant hence broadband interferometric gravitational-wave detection. First, the factor $1/K$ stems from the resonant mechanical strain-to-twist enhancement discussed in Fig.\,2. This mechanism enhances the gravitational-wave induced strain, environmental noise, and thermal fluctuations alike. Thermal fluctuation can be estimated from  Eq.\,(16) in \cite{Saulson1990thermal}. For example, for the parameters given in Fig.\,1, we have the mechanical quality factor $1/\mathrm{tan}(\delta) \approx 3.3\times10^{5}$, the axial stiffness $3.33 \times 10^{10}\,\mathrm{N/m}$, the mass $2.4\,\mathrm{kg}$, the height $a + b = 8\,\mathrm{cm}$ and the eigenfrequency $10.4\,\mathrm{kHz}$, leading to a strain of $\approx10^{-21}/\sqrt{\rm{Hz}}$ at a temperature of $40\,\mathrm{mK}$. Other sources of thermodynamic noise \cite{Braginsky:1999rp} are subdominant. The device needs to be suitably decoupled from environmental noise. Furthermore, reducing technical noise on the angle of the incident laser beam is critical. Second, the factor $\lambda/(Nw)$ can be seen as a lever-arm effect of the optical cavity resulting from the angle detection scheme. For the  parameters considered above, the accessible number of round trips in the cavity $N$ is essentially limited by diffraction of light. In the regime where the maximum $N$ is rather determined by the dwell-time bound (5), the strain sensitivity of our setup scales $\propto 1/(Nw)$, whereas that of a linear interferometer scales $\propto 1/(NL)$, and thus our setup could potentially lead to a sensitivity gain $\propto L/w$. 

As a conservative example, the parameters $\lambda = 1.064\,{\rm \mu m}$, $N = 2 \times 10^4$, $w = 5\,{\rm mm}$, $K = 100$ (compare Fig.\,2 with peak enhancement $>10^5$), and $R = 10^{20}\,{\rm s}^{-1}$, hence $P = R \hbar c_0 2\pi / \lambda = 18.6\,{\rm W}$ optical power, lead to the gravitational-wave strain-sensitivity estimate of
\begin{equation} \label{(10)} 
h_0 = 1.2 \times 10^{-21}/\sqrt{\rm Hz}
\end{equation}
at around $10\,\rm kHz$ frequency. From above, we additionally have $L = 1\,{\rm mm}$ and $D = 10 w = 5\,{\rm cm}$ (cf. Fig.\,1). To scale our detector to ten times higher (lower) frequencies, the geometrical parameters shown in Fig.\,1 and the beam radius $w$ can be decreased (increased) tenfold, while adjusting the product $NL\propto w^2$ according to (6). In this case, for fixed $N$, the strain sensitivity scales proportional to frequency. Towards lower frequencies, the sensitivity could alternatively be improved by fixing $w$ and increasing $N$.

In conclusion, we have proposed a novel compact approach for detecting gravitational waves at frequencies in the range of $1$ -- $100\,{\rm kHz}$. A resonant chiral mechanical element converts a gravitational-wave strain into the rotation of one flat end mirror of a short unstable optical Fabry-Perot cavity. The unstable optical cavity enhances the deflection angle by multiple bounces of light within. Important parameters include the laser power $P$, the mean number of round trips in the cavity $N$, and the mechanical enhancement factor $K$. For $P = 18.6\, {\rm W}$ laser power at  $\lambda = 1.064\,{\rm \mu m}$ wavelength, $w = 5\,{\rm mm}$ beam radius, $N = 2 \times 10^4$, and $K = 100$, we have estimated a gravitational-wave strain sensitivity of better than $h_0 = 1.2 \times 10^{-21} / \sqrt{\rm Hz}$ in a bandwidth of about $100\,{\rm Hz}$ at around $10\,{\rm kHz}$ frequency. Using somewhat more optimistic parameters of, e.g., $P = 100\, {\rm W}$, $N = 4 \times 10^4$, and $K = 1000$, the theoretical sensitivity estimate improves to $h_0 = 2.6 \times 10^{-23} / \sqrt{\rm Hz}$. Using a smaller laser wavelength $\lambda$ would further improve the behavior. For comparison, LIGO has achieved experimentally a squeezed-states-of-light enhanced sensitivity of about $h_0 = 4.5 \times 10^{-23} / \sqrt{\rm Hz}$ at around $10\,\rm kHz$ frequency \cite{aasi2013enhanced}.  

\section{Supplemental Materials}
\subsection{Eigenmodes of the chiral mechanical system}

See figure \ref{FigureS1}.

\begin{figure}
\includegraphics[width=9.0cm,angle=0]{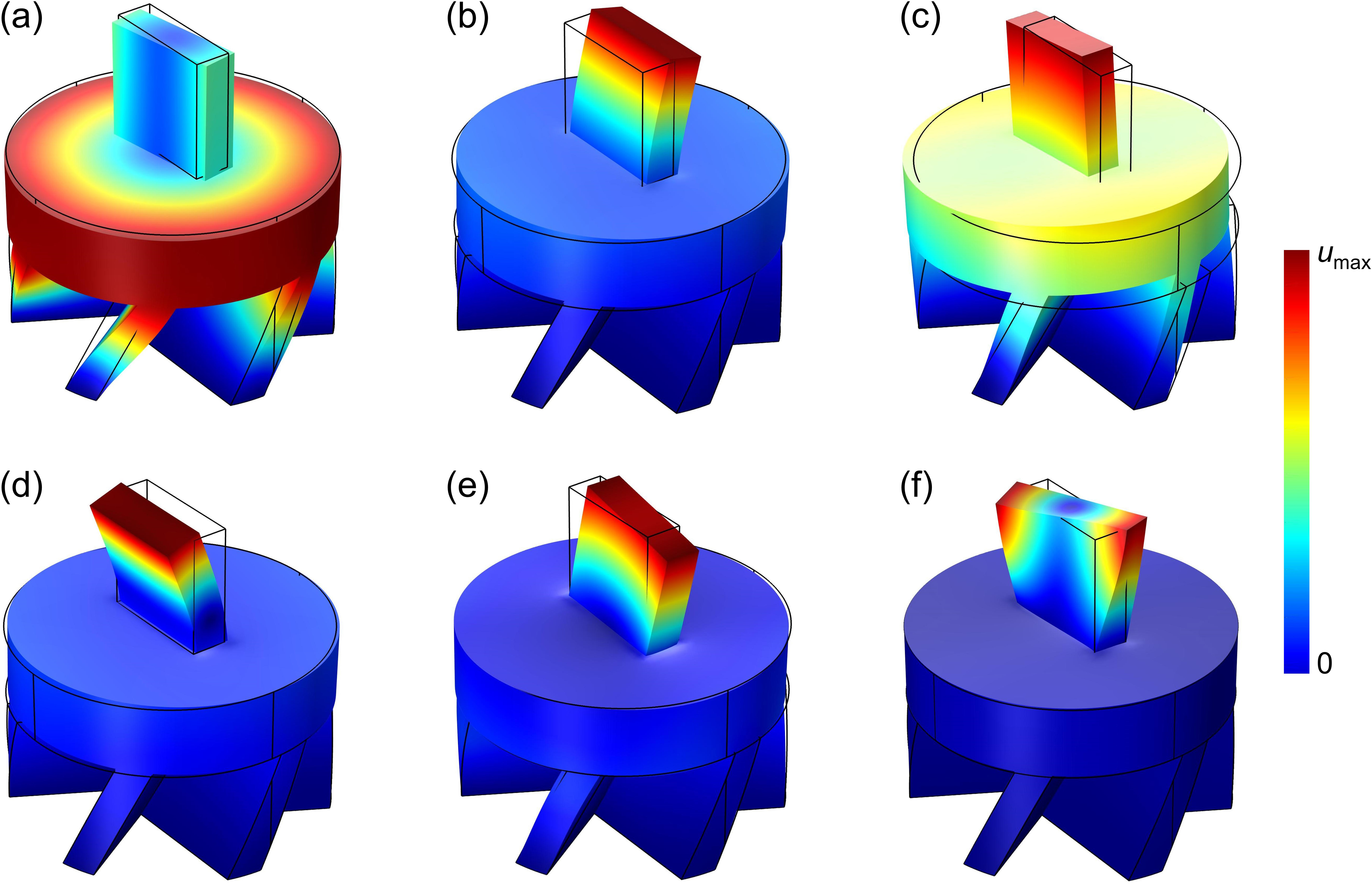}
\caption{Six lowest-frequency eigenmodes of the system shown in Fig.\,1. The lowest mechanical eigenmode of the overall setup corresponds to the torsional eigenmode. The eigenfrequencies are (a) $10.40\,{\rm kHz}$, (b) $11.29\,{\rm kHz}$, (c) $12.25\,{\rm kHz}$, (d) $16.10\,{\rm kHz}$, (e) $29.86\,{\rm kHz}$ and (f) $32.45\,{\rm kHz}$. }
\label{FigureS1}
\end{figure}

\begin{figure}
\includegraphics[width=9.0cm,angle=0]{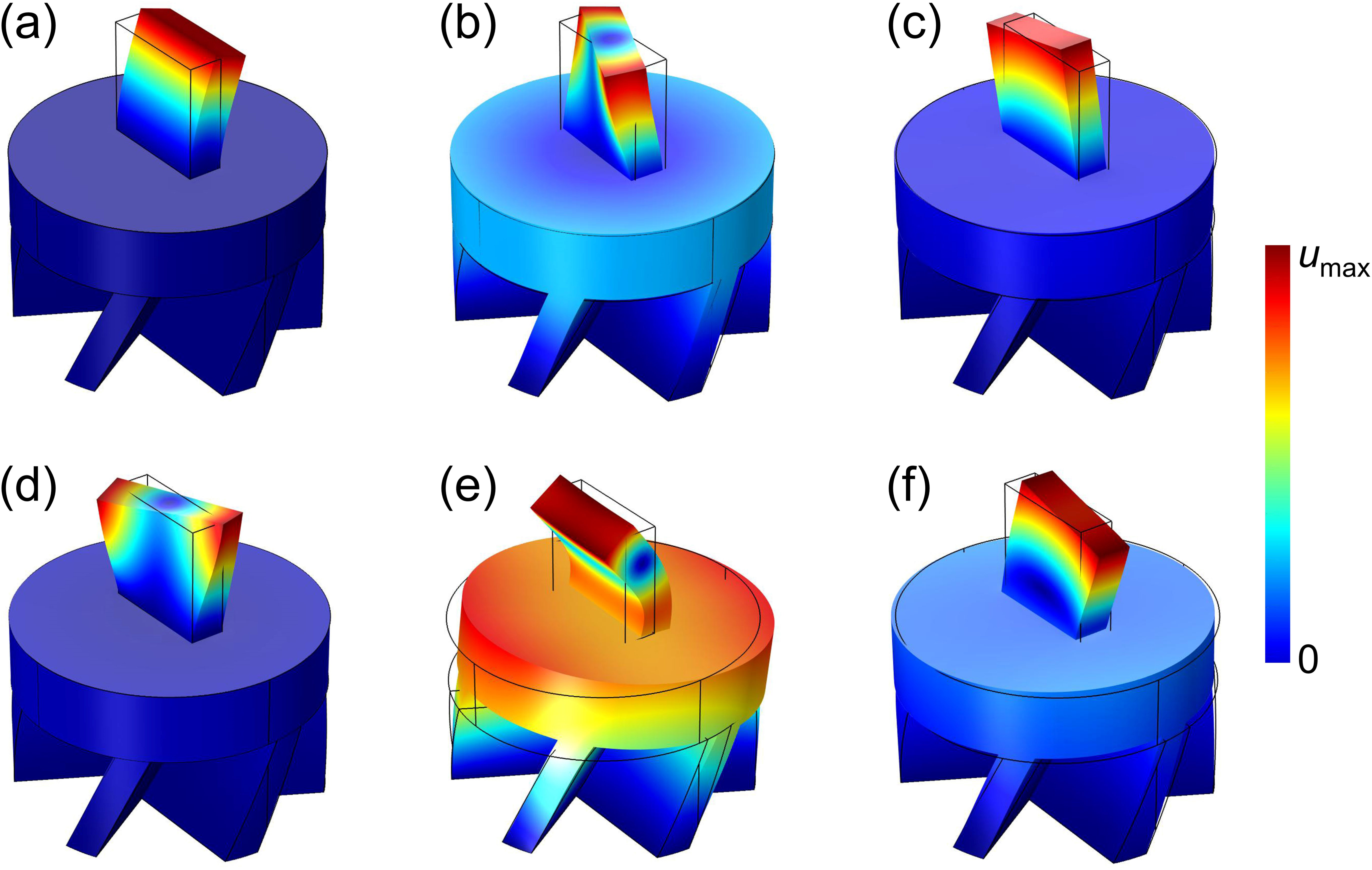}
\caption{Six lowest-frequency eigenmodes of the system with diamond bottom and silica mirror substrate. The second mechanical eigenmode corresponds to the torsional eigenmode, and the mirror substrate warps due to its low stiffness. The eigenfrequencies are (a) $5.10\,{\rm kHz}$, (b) $10.31\,{\rm kHz}$, (c) $10.87\,{\rm kHz}$, (d) $11.32\,{\rm kHz}$, (e) $13.21\,{\rm kHz}$ and (f) $14.40\,{\rm kHz}$. }
\label{FigureS1}
\end{figure}

\subsection{Calculated response of the chiral mechanical system for other material combinations}
\begin{figure*}[t]
\includegraphics[width=14.4cm,angle=0]{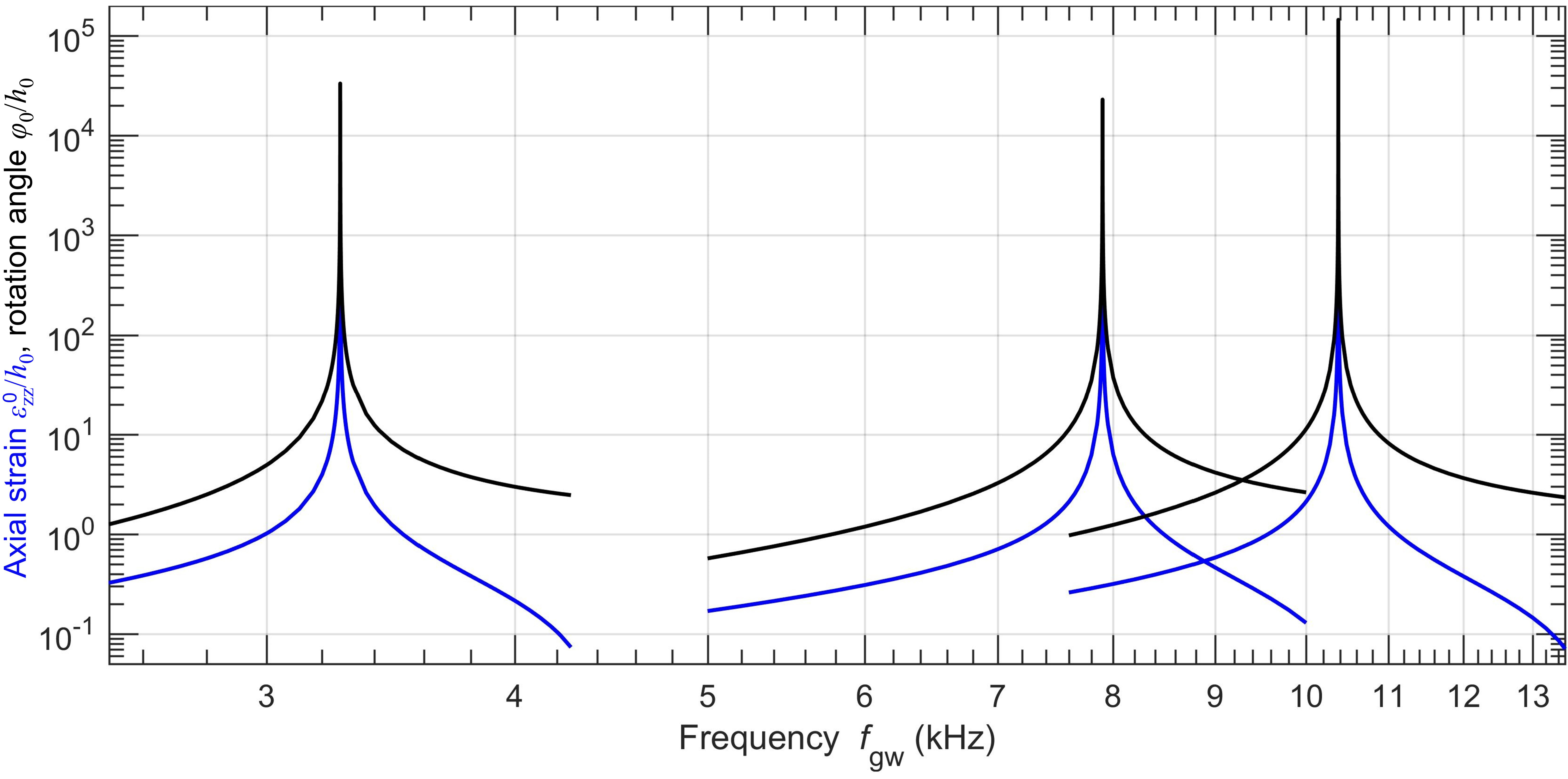}
\caption{Calculations as in Fig.\,2, but for other material combinations. Left: silica bottom, sapphire mirror substrate. Middle: boron-carbide bottom, sapphire mirror substrate. Right: diamond bottom, sapphire mirror substrate. As in the main text, we also choose an isotropic (polycrystalline) version for the materials silica, sapphire and boron-carbide [17]. The material parameters of silica are: Young’s modulus of $E = 70\,{\rm GPa}$, loss tangent ${\rm tan}(\delta) = 1.4 \times 10^{-5}$, Poisson’s ratio $\nu = 0.17$, and mass density $\rho = 2.2 \times 10^3\,{\rm kg/m^3}$. The material parameters of sapphire are: Young’s modulus $E = 445\,{\rm GPa}$, loss tangent ${\rm tan}(\delta) = 7.5 \times 10^{-6}$, Poisson’s ratio $\nu = 0.24$, and mass density of $\rho = 3.98 \times 10^3\,{\rm kg/m^3}$. The material parameters of boron-carbide are: Young’s modulus $E = 450\,{\rm GPa}$, loss tangent ${\rm tan}(\delta) = 2 \times 10^{-5}$, Poisson’s ratio $\nu = 0.2$, and mass density $\rho = 2.4 \times 10^3\,{\rm kg/m^3}$ . }
\label{FigureS2}
\end{figure*}

See figure \ref{FigureS2}.

\section{Acknowledgments}
We acknowledge discussions with Dima Budker (University of Mainz), Jason Hogan (Stanford University) as well as with Andreas Naber, Tobias Frenzel, and Julian K\"opfler (all KIT). Y.C. acknowledges support by the Alexander von Humboldt foundation and by the National Natural Science Foundation of China (contract No. 11802017). This research has additionally been funded by the Deutsche Forschungsgemeinschaft (DFG, German Research Foundation) under Germany's Excellence Strategy {\it via} the Excellence Cluster “3D Matter Made to Order” (EXC-2082/1-390761711), which has also been supported by the Carl Zeiss Foundation through the “Carl-Zeiss-Foundation-Focus@HEiKA”, by the State of Baden-Württemberg, and by the Karlsruhe Institute of Technology (KIT). We further acknowledge support by the Helmholtz program “Science and Technology of Nanosystems” (STN), and by the associated KIT project “Virtual Materials Design” (VIRTMAT). M.K. is grateful for support by the EIPHI Graduate School (contract No. ANR-17-EURE-0002) and by the French Investissements d'Avenir program, project ISITEBFC (contract No. ANR-15-IDEX-03). D.E.K. and S.R. are supported by the US National Science Foundation (contract No. PHY-1818899). S.R. is also supported by the DoE under a  QuantISED grant for MAGIS. A.O.S. is supported by the US National Science Foundation (contract No. 1806557), US Department of Energy (contract No. DE-SC0019450), and the Simons Foundation (contract No. 641332).


\begin{thebibliography}{23}%
\makeatletter
\providecommand \@ifxundefined [1]{%
 \@ifx{#1\undefined}
}%
\providecommand \@ifnum [1]{%
 \ifnum #1\expandafter \@firstoftwo
 \else \expandafter \@secondoftwo
 \fi
}%
\providecommand \@ifx [1]{%
 \ifx #1\expandafter \@firstoftwo
 \else \expandafter \@secondoftwo
 \fi
}%
\providecommand \natexlab [1]{#1}%
\providecommand \enquote  [1]{``#1''}%
\providecommand \bibnamefont  [1]{#1}%
\providecommand \bibfnamefont [1]{#1}%
\providecommand \citenamefont [1]{#1}%
\providecommand \href@noop [0]{\@secondoftwo}%
\providecommand \href [0]{\begingroup \@sanitize@url \@href}%
\providecommand \@href[1]{\@@startlink{#1}\@@href}%
\providecommand \@@href[1]{\endgroup#1\@@endlink}%
\providecommand \@sanitize@url [0]{\catcode `\\12\catcode `\$12\catcode
  `\&12\catcode `\#12\catcode `\^12\catcode `\_12\catcode `\%12\relax}%
\providecommand \@@startlink[1]{}%
\providecommand \@@endlink[0]{}%
\providecommand \url  [0]{\begingroup\@sanitize@url \@url }%
\providecommand \@url [1]{\endgroup\@href {#1}{\urlprefix }}%
\providecommand \urlprefix  [0]{URL }%
\providecommand \Eprint [0]{\href }%
\providecommand \doibase [0]{http://dx.doi.org/}%
\providecommand \selectlanguage [0]{\@gobble}%
\providecommand \bibinfo  [0]{\@secondoftwo}%
\providecommand \bibfield  [0]{\@secondoftwo}%
\providecommand \translation [1]{[#1]}%
\providecommand \BibitemOpen [0]{}%
\providecommand \bibitemStop [0]{}%
\providecommand \bibitemNoStop [0]{.\EOS\space}%
\providecommand \EOS [0]{\spacefactor3000\relax}%
\providecommand \BibitemShut  [1]{\csname bibitem#1\endcsname}%
\let\auto@bib@innerbib\@empty
\bibitem [{\citenamefont {Einstein}(1916)}]{einstein1916approximative}%
  \BibitemOpen
  \bibfield  {author} {\bibinfo {author} {\bibfnamefont {A.}~\bibnamefont
  {Einstein}},\ }\href@noop {} {\bibfield  {journal} {\bibinfo  {journal}
  {Sitzungsber. Preuss. Akad. Wiss. Berlin (Math. Phys.)}\ }\textbf {\bibinfo
  {volume} {1916}},\ \bibinfo {pages} {688} (\bibinfo {year}
  {1916})}\BibitemShut {NoStop}%
\bibitem [{\citenamefont {Abbott}\ \emph
  {et~al.}(2016{\natexlab{a}})\citenamefont {Abbott} \emph
  {et~al.}}]{Abbott2016Obser}%
  \BibitemOpen
  \bibfield  {author} {\bibinfo {author} {\bibfnamefont {B.~P.}\ \bibnamefont
  {Abbott}} \emph {et~al.} (\bibinfo {collaboration} {LIGO Scientific
  Collaboration and Virgo Collaboration}),\ }\href {\doibase
  10.1103/PhysRevLett.116.061102} {\bibfield  {journal} {\bibinfo  {journal}
  {Phys. Rev. Lett.}\ }\textbf {\bibinfo {volume} {116}},\ \bibinfo {pages}
  {061102} (\bibinfo {year} {2016}{\natexlab{a}})}\BibitemShut {NoStop}%
\bibitem [{\citenamefont {Abbott}\ \emph
  {et~al.}(2016{\natexlab{b}})\citenamefont {Abbott} \emph
  {et~al.}}]{GW1512262016Abbott}%
  \BibitemOpen
  \bibfield  {author} {\bibinfo {author} {\bibfnamefont {B.~P.}\ \bibnamefont
  {Abbott}} \emph {et~al.} (\bibinfo {collaboration} {LIGO Scientific
  Collaboration and Virgo Collaboration}),\ }\href {\doibase
  10.1103/PhysRevLett.116.241103} {\bibfield  {journal} {\bibinfo  {journal}
  {Phys. Rev. Lett.}\ }\textbf {\bibinfo {volume} {116}},\ \bibinfo {pages}
  {241103} (\bibinfo {year} {2016}{\natexlab{b}})}\BibitemShut {NoStop}%
\bibitem [{\citenamefont {Abbott}\ \emph
  {et~al.}(2017{\natexlab{a}})\citenamefont {Abbott} \emph
  {et~al.}}]{GW1708172017Abbott}%
  \BibitemOpen
  \bibfield  {author} {\bibinfo {author} {\bibfnamefont {B.~P.}\ \bibnamefont
  {Abbott}} \emph {et~al.} (\bibinfo {collaboration} {LIGO Scientific
  Collaboration and Virgo Collaboration}),\ }\href {\doibase
  10.1103/PhysRevLett.119.161101} {\bibfield  {journal} {\bibinfo  {journal}
  {Phys. Rev. Lett.}\ }\textbf {\bibinfo {volume} {119}},\ \bibinfo {pages}
  {161101} (\bibinfo {year} {2017}{\natexlab{a}})}\BibitemShut {NoStop}%
\bibitem [{\citenamefont {Abbott}\ \emph
  {et~al.}(2017{\natexlab{b}})\citenamefont {Abbott} \emph
  {et~al.}}]{GW1701042017Abbott}%
  \BibitemOpen
  \bibfield  {author} {\bibinfo {author} {\bibfnamefont {B.~P.}\ \bibnamefont
  {Abbott}} \emph {et~al.} (\bibinfo {collaboration} {LIGO Scientific and Virgo
  Collaboration}),\ }\href {\doibase 10.1103/PhysRevLett.118.221101} {\bibfield
   {journal} {\bibinfo  {journal} {Phys. Rev. Lett.}\ }\textbf {\bibinfo
  {volume} {118}},\ \bibinfo {pages} {221101} (\bibinfo {year}
  {2017}{\natexlab{b}})}\BibitemShut {NoStop}%
\bibitem [{\citenamefont {Aasi}\ \emph {et~al.}(2015)\citenamefont {Aasi} \emph
  {et~al.}}]{aasi2015advanced}%
  \BibitemOpen
  \bibfield  {author} {\bibinfo {author} {\bibfnamefont {J.}~\bibnamefont
  {Aasi}} \emph {et~al.} (\bibinfo {collaboration} {LIGO Scientific and Virgo
  Collaboration}),\ }\href {\doibase 10.1088/0264-9381/32/7/074001} {\bibfield
  {journal} {\bibinfo  {journal} {Classical and quantum gravity}\ }\textbf
  {\bibinfo {volume} {32}},\ \bibinfo {pages} {074001} (\bibinfo {year}
  {2015})}\BibitemShut {NoStop}%
\bibitem [{\citenamefont {Kaplan}\ and\ \citenamefont
  {Rajendran}(2019)}]{Kaplan:2018dqx}%
  \BibitemOpen
  \bibfield  {author} {\bibinfo {author} {\bibfnamefont {D.~E.}\ \bibnamefont
  {Kaplan}}\ and\ \bibinfo {author} {\bibfnamefont {S.}~\bibnamefont
  {Rajendran}},\ }\href {\doibase 10.1103/PhysRevD.99.044033} {\bibfield
  {journal} {\bibinfo  {journal} {Phys. Rev. D}\ }\textbf {\bibinfo {volume}
  {99}},\ \bibinfo {pages} {044033} (\bibinfo {year} {2019})}\BibitemShut
  {NoStop}%
\bibitem [{\citenamefont {Arvanitaki}\ \emph {et~al.}(2010)\citenamefont
  {Arvanitaki}, \citenamefont {Dimopoulos}, \citenamefont {Dubovsky},
  \citenamefont {Kaloper},\ and\ \citenamefont
  {March-Russell}}]{Arvanitaki:2009fg}%
  \BibitemOpen
  \bibfield  {author} {\bibinfo {author} {\bibfnamefont {A.}~\bibnamefont
  {Arvanitaki}}, \bibinfo {author} {\bibfnamefont {S.}~\bibnamefont
  {Dimopoulos}}, \bibinfo {author} {\bibfnamefont {S.}~\bibnamefont
  {Dubovsky}}, \bibinfo {author} {\bibfnamefont {N.}~\bibnamefont {Kaloper}}, \
  and\ \bibinfo {author} {\bibfnamefont {J.}~\bibnamefont {March-Russell}},\
  }\href {\doibase 10.1103/PhysRevD.81.123530} {\bibfield  {journal} {\bibinfo
  {journal} {Phys. Rev. D}\ }\textbf {\bibinfo {volume} {81}},\ \bibinfo
  {pages} {123530} (\bibinfo {year} {2010})}\BibitemShut {NoStop}%
\bibitem [{\citenamefont {Ranjit}\ \emph {et~al.}(2016)\citenamefont {Ranjit},
  \citenamefont {Cunningham}, \citenamefont {Casey},\ and\ \citenamefont
  {Geraci}}]{Gambhir2016Zeptonewton}%
  \BibitemOpen
  \bibfield  {author} {\bibinfo {author} {\bibfnamefont {G.}~\bibnamefont
  {Ranjit}}, \bibinfo {author} {\bibfnamefont {M.}~\bibnamefont {Cunningham}},
  \bibinfo {author} {\bibfnamefont {K.}~\bibnamefont {Casey}}, \ and\ \bibinfo
  {author} {\bibfnamefont {A.~A.}\ \bibnamefont {Geraci}},\ }\href {\doibase
  10.1103/PhysRevA.93.053801} {\bibfield  {journal} {\bibinfo  {journal} {Phys.
  Rev. A}\ }\textbf {\bibinfo {volume} {93}},\ \bibinfo {pages} {053801}
  (\bibinfo {year} {2016})}\BibitemShut {NoStop}%
\bibitem [{\citenamefont {Graham}\ \emph {et~al.}(2016)\citenamefont {Graham},
  \citenamefont {Kaplan}, \citenamefont {Mardon}, \citenamefont {Rajendran},\
  and\ \citenamefont {Terrano}}]{Graham:2015ifn}%
  \BibitemOpen
  \bibfield  {author} {\bibinfo {author} {\bibfnamefont {P.~W.}\ \bibnamefont
  {Graham}}, \bibinfo {author} {\bibfnamefont {D.~E.}\ \bibnamefont {Kaplan}},
  \bibinfo {author} {\bibfnamefont {J.}~\bibnamefont {Mardon}}, \bibinfo
  {author} {\bibfnamefont {S.}~\bibnamefont {Rajendran}}, \ and\ \bibinfo
  {author} {\bibfnamefont {W.~A.}\ \bibnamefont {Terrano}},\ }\href {\doibase
  10.1103/PhysRevD.93.075029} {\bibfield  {journal} {\bibinfo  {journal} {Phys.
  Rev. D}\ }\textbf {\bibinfo {volume} {93}},\ \bibinfo {pages} {075029}
  (\bibinfo {year} {2016})}\BibitemShut {NoStop}%
\bibitem [{\citenamefont {Arvanitaki}\ \emph {et~al.}(2018)\citenamefont
  {Arvanitaki}, \citenamefont {Graham}, \citenamefont {Hogan}, \citenamefont
  {Rajendran},\ and\ \citenamefont {Van~Tilburg}}]{Arvanitaki:2016fyj}%
  \BibitemOpen
  \bibfield  {author} {\bibinfo {author} {\bibfnamefont {A.}~\bibnamefont
  {Arvanitaki}}, \bibinfo {author} {\bibfnamefont {P.~W.}\ \bibnamefont
  {Graham}}, \bibinfo {author} {\bibfnamefont {J.~M.}\ \bibnamefont {Hogan}},
  \bibinfo {author} {\bibfnamefont {S.}~\bibnamefont {Rajendran}}, \ and\
  \bibinfo {author} {\bibfnamefont {K.}~\bibnamefont {Van~Tilburg}},\ }\href
  {\doibase 10.1103/PhysRevD.97.075020} {\bibfield  {journal} {\bibinfo
  {journal} {Phys. Rev. D}\ }\textbf {\bibinfo {volume} {97}},\ \bibinfo
  {pages} {075020} (\bibinfo {year} {2018})}\BibitemShut {NoStop}%
\bibitem [{\citenamefont {Frenzel}\ \emph {et~al.}(2017)\citenamefont
  {Frenzel}, \citenamefont {Kadic},\ and\ \citenamefont
  {Wegener}}]{Frenzel2017}%
  \BibitemOpen
  \bibfield  {author} {\bibinfo {author} {\bibfnamefont {T.}~\bibnamefont
  {Frenzel}}, \bibinfo {author} {\bibfnamefont {M.}~\bibnamefont {Kadic}}, \
  and\ \bibinfo {author} {\bibfnamefont {M.}~\bibnamefont {Wegener}},\ }\href
  {\doibase 10.1126/science.aao4640} {\bibfield  {journal} {\bibinfo  {journal}
  {Science}\ }\textbf {\bibinfo {volume} {358}},\ \bibinfo {pages} {1072}
  (\bibinfo {year} {2017})}\BibitemShut {NoStop}%
\bibitem [{\citenamefont {Fernandez-Corbaton}\ \emph
  {et~al.}(2019)\citenamefont {Fernandez-Corbaton}, \citenamefont {Rockstuhl},
  \citenamefont {Ziemke}, \citenamefont {Gumbsch}, \citenamefont {Albiez},
  \citenamefont {Schwaiger}, \citenamefont {Frenzel}, \citenamefont {Kadic},\
  and\ \citenamefont {Wegener}}]{fernandez2019new}%
  \BibitemOpen
  \bibfield  {author} {\bibinfo {author} {\bibfnamefont {I.}~\bibnamefont
  {Fernandez-Corbaton}}, \bibinfo {author} {\bibfnamefont {C.}~\bibnamefont
  {Rockstuhl}}, \bibinfo {author} {\bibfnamefont {P.}~\bibnamefont {Ziemke}},
  \bibinfo {author} {\bibfnamefont {P.}~\bibnamefont {Gumbsch}}, \bibinfo
  {author} {\bibfnamefont {A.}~\bibnamefont {Albiez}}, \bibinfo {author}
  {\bibfnamefont {R.}~\bibnamefont {Schwaiger}}, \bibinfo {author}
  {\bibfnamefont {T.}~\bibnamefont {Frenzel}}, \bibinfo {author} {\bibfnamefont
  {M.}~\bibnamefont {Kadic}}, \ and\ \bibinfo {author} {\bibfnamefont
  {M.}~\bibnamefont {Wegener}},\ }\href {\doibase 10.1002/adma.201807742}
  {\bibfield  {journal} {\bibinfo  {journal} {Adv. Mater.}\ }\textbf {\bibinfo
  {volume} {31}},\ \bibinfo {pages} {1807742} (\bibinfo {year}
  {2019})}\BibitemShut {NoStop}%
\bibitem [{\citenamefont {Chen}\ \emph {et~al.}(2020)\citenamefont {Chen},
  \citenamefont {Frenzel}, \citenamefont {Guenneau}, \citenamefont {Kadic},\
  and\ \citenamefont {Wegener}}]{chen2020mapping}%
  \BibitemOpen
  \bibfield  {author} {\bibinfo {author} {\bibfnamefont {Y.}~\bibnamefont
  {Chen}}, \bibinfo {author} {\bibfnamefont {T.}~\bibnamefont {Frenzel}},
  \bibinfo {author} {\bibfnamefont {S.}~\bibnamefont {Guenneau}}, \bibinfo
  {author} {\bibfnamefont {M.}~\bibnamefont {Kadic}}, \ and\ \bibinfo {author}
  {\bibfnamefont {M.}~\bibnamefont {Wegener}},\ }\href {\doibase
  10.1016/j.jmps.2020.103877} {\bibfield  {journal} {\bibinfo  {journal} {J.
  Mech. Phys. Solids}\ }\textbf {\bibinfo {volume} {137}},\ \bibinfo {pages}
  {103877} (\bibinfo {year} {2020})}\BibitemShut {NoStop}%
\bibitem [{\citenamefont {Hogan}\ \emph {et~al.}(2011)\citenamefont {Hogan},
  \citenamefont {Hammer}, \citenamefont {Chiow}, \citenamefont {Dickerson},
  \citenamefont {Johnson}, \citenamefont {Kovachy}, \citenamefont
  {Sugarbaker},\ and\ \citenamefont {Kasevich}}]{Hogan_2011}%
  \BibitemOpen
  \bibfield  {author} {\bibinfo {author} {\bibfnamefont {J.~M.}\ \bibnamefont
  {Hogan}}, \bibinfo {author} {\bibfnamefont {J.}~\bibnamefont {Hammer}},
  \bibinfo {author} {\bibfnamefont {S.-W.}\ \bibnamefont {Chiow}}, \bibinfo
  {author} {\bibfnamefont {S.}~\bibnamefont {Dickerson}}, \bibinfo {author}
  {\bibfnamefont {D.~M.~S.}\ \bibnamefont {Johnson}}, \bibinfo {author}
  {\bibfnamefont {T.}~\bibnamefont {Kovachy}}, \bibinfo {author} {\bibfnamefont
  {A.}~\bibnamefont {Sugarbaker}}, \ and\ \bibinfo {author} {\bibfnamefont
  {M.~A.}\ \bibnamefont {Kasevich}},\ }\href {\doibase 10.1364/ol.36.001698}
  {\bibfield  {journal} {\bibinfo  {journal} {Opt. Lett.}\ }\textbf {\bibinfo
  {volume} {36}},\ \bibinfo {pages} {1698} (\bibinfo {year}
  {2011})}\BibitemShut {NoStop}%
\bibitem [{\citenamefont {Misner}\ \emph {et~al.}(1973)\citenamefont {Misner},
  \citenamefont {Thorne},\ and\ \citenamefont
  {Wheeler}}]{misner1973gravitation}%
  \BibitemOpen
  \bibfield  {author} {\bibinfo {author} {\bibfnamefont {C.~W.}\ \bibnamefont
  {Misner}}, \bibinfo {author} {\bibfnamefont {K.~S.}\ \bibnamefont {Thorne}},
  \ and\ \bibinfo {author} {\bibfnamefont {J.~A.}\ \bibnamefont {Wheeler}},\
  }\href {\doibase 10.1086/664983} {\emph {\bibinfo {title} {Gravitation}}}\
  (\bibinfo  {publisher} {Princeton University Press},\ \bibinfo {year}
  {1973})\BibitemShut {NoStop}%
\bibitem [{\citenamefont {Hughes}(2000)}]{thomas2000finite}%
  \BibitemOpen
  \bibfield  {author} {\bibinfo {author} {\bibfnamefont {T.~J.}\ \bibnamefont
  {Hughes}},\ }\href {\doibase 10.1111/j.1467-8667.1989.tb00025.x} {\emph
  {\bibinfo {title} {The Finite Element Method}}}\ (\bibinfo  {publisher}
  {Dover Publications, New York},\ \bibinfo {year} {2000})\BibitemShut
  {NoStop}%
\bibitem [{\citenamefont {Ashby}(2011)}]{Ashby2011}%
  \BibitemOpen
  \bibfield  {author} {\bibinfo {author} {\bibfnamefont {M.~F.}\ \bibnamefont
  {Ashby}},\ }\href {\doibase 10.1016/B978-1-85617-663-7.00001-1} {\emph
  {\bibinfo {title} {Materials Selection in Mechanical Design}}}\ (\bibinfo
  {publisher} {Butterworth-Heinemann, Oxford},\ \bibinfo {year}
  {2011})\BibitemShut {NoStop}%
\bibitem [{\citenamefont {Koechner}(2006)}]{koechner2013solid}%
  \BibitemOpen
  \bibfield  {author} {\bibinfo {author} {\bibfnamefont {W.}~\bibnamefont
  {Koechner}},\ }\href {\doibase 10.1007/0-387-29338-8} {\emph {\bibinfo
  {title} {Solid-State Laser Engineering}}},\ Vol.~\bibinfo {volume} {1}\
  (\bibinfo  {publisher} {Springer, New York},\ \bibinfo {year}
  {2006})\BibitemShut {NoStop}%
\bibitem [{\citenamefont {Saleh}\ and\ \citenamefont
  {Teich}(2019)}]{Fundamentals2019Bahaa}%
  \BibitemOpen
  \bibfield  {author} {\bibinfo {author} {\bibfnamefont {B.~E.~A.}\
  \bibnamefont {Saleh}}\ and\ \bibinfo {author} {\bibfnamefont {M.~C.}\
  \bibnamefont {Teich}},\ }\href {\doibase 10.1002/0471213748} {\emph {\bibinfo
  {title} {Fundamentals of Photonics}}},\ Vol.~\bibinfo {volume} {2}\ (\bibinfo
   {publisher} {Wiley, New York},\ \bibinfo {year} {2019})\BibitemShut
  {NoStop}%
\bibitem [{\citenamefont {Saulson}(1990)}]{Saulson1990thermal}%
  \BibitemOpen
  \bibfield  {author} {\bibinfo {author} {\bibfnamefont {P.~R.}\ \bibnamefont
  {Saulson}},\ }\href {\doibase 10.1103/PhysRevD.42.2437} {\bibfield  {journal}
  {\bibinfo  {journal} {Phys. Rev. D}\ }\textbf {\bibinfo {volume} {42}},\
  \bibinfo {pages} {2437} (\bibinfo {year} {1990})}\BibitemShut {NoStop}%
\bibitem [{\citenamefont {Braginsky}\ \emph {et~al.}(1999)\citenamefont
  {Braginsky}, \citenamefont {Gorodetsky},\ and\ \citenamefont
  {Vyatchanin}}]{Braginsky:1999rp}%
  \BibitemOpen
  \bibfield  {author} {\bibinfo {author} {\bibfnamefont {V.}~\bibnamefont
  {Braginsky}}, \bibinfo {author} {\bibfnamefont {M.}~\bibnamefont
  {Gorodetsky}}, \ and\ \bibinfo {author} {\bibfnamefont {S.}~\bibnamefont
  {Vyatchanin}},\ }\href {\doibase 10.1016/S0375-9601(99)00785-9} {\bibfield
  {journal} {\bibinfo  {journal} {Phys. Lett. A}\ }\textbf {\bibinfo {volume}
  {264}},\ \bibinfo {pages} {1} (\bibinfo {year} {1999})}\BibitemShut {NoStop}%
\bibitem [{\citenamefont {Aasi}\ \emph {et~al.}(2013)\citenamefont {Aasi} \emph
  {et~al.}}]{aasi2013enhanced}%
  \BibitemOpen
  \bibfield  {author} {\bibinfo {author} {\bibfnamefont {J.}~\bibnamefont
  {Aasi}} \emph {et~al.} (\bibinfo {collaboration} {LIGO Scientific and Virgo
  Collaboration}),\ }\href {\doibase 10.1038/nphoton.2013.177} {\bibfield
  {journal} {\bibinfo  {journal} {Nature Photonics}\ }\textbf {\bibinfo
  {volume} {7}},\ \bibinfo {pages} {613} (\bibinfo {year} {2013})}\BibitemShut
  {NoStop}%
 \bibitem{PRLSupp} See Supplemental Material at [URL will be inserted by publisher] for eigenmodes of the mechanical chiral element and calculated response for other material combinations.%
\end{thebibliography}

%

\end{document}